\documentclass[twocolumn,showpacs,aps]{revtex4}
\usepackage{graphicx}
 \pdfoutput=1

\begin{document}
\preprint{PRA}
\title{{\it p}-wave stabilization of three-dimensional Bose-Fermi solitons}

\author{N G Parker}
\affiliation{School of Mathematics and Statistics, Newcastle
University, Newcastle Upon Tyne, UK}
\email{nick.parker@ncl.ac.uk}
\author{D A Smith}
\affiliation{Vienna Center for Quantum Science and Technology, Atominstitut, TU Wien,
Vienna, Austria}

\begin{abstract}
We explore bright soliton solutions of ultracold Bose-Fermi gases, showing that the presence of {\it p}-wave interactions can remove the usual collapse instability and support stable soliton solutions that are global energy minima.  A variational model that incorporates the relevant {\it s}- and {\it p}-wave interactions in the system is established analytically and solved numerically to probe the dependencies of the solitons on key experimental parameters.  Under attractive {\it s}-wave interactions, bright solitons exist only as meta-stable states susceptible to collapse.  Remarkably, the presence of repulsive {\it p}-wave interactions alleviates this collapse instability.  This dramatically widens the range of experimentally-achievable soliton solutions and indicates greatly enhanced robustness.  While we focus specifically on the boson-fermion pairing of $^{87}$Rb and $^{40}$K, the stabilization inferred by repulsive {\it p}-wave interactions should apply to the wider remit of ultracold Bose-Fermi mixtures.

\end{abstract}
\date{\today}
\maketitle

\section{Introduction}
The study of wave mechanics and propagation in non-linear media is a
fundamental concept within physics.  In particular, solitons are a
general non-dispersive solution to the one-dimensional non-linear
wave equation.  Bright solitons have been observed in many areas of
physics, such as in water \cite{ScottRussell}, liquid hydrogen
\cite{hydrogen}, optics \cite{optmat} and atomic Bose-Einstein
condensates (BECs) \cite{Khaykovich,Strecker,Cornish}.  In the latter case,
these self-trapped matter waves are supported by a balance between
attractive atomic scattering interactions and repulsive zero-point kinetic
energy between bosons.  As well as being of fundamental interest in many-body
quantum physics, bright matter-wave solitons are being touted for potential uses in atom interferometry \cite{Khaykovich,Strecker,Parker,Billam} and surface characterization
\cite{CornishReflect2008}.  However, when realized in
three-dimensions these solitons exist only as meta-stable states
prone to catastrophic collapse \cite{Gaussian_solitons,carr}.

Ultracold Bose-Fermi (BF) gases have received a great deal of recent
experimental attention and have been realised through $^{7}$Li-$^{6}$Li \cite{Li6-Li7},
$^{23}$Na-$^{6}$Li \cite{Na-Li}, $^{87}$Rb-$^{40}$K \cite{Rb-K},
$^{174}$Yb-$^{173}$Yb \cite{Fukuhara2009} and $^{84}$Sr-$^{87}$Sr \cite{Tey2010}
mixtures.  At such low temperatures, the scattering of atoms with non-zero relative angular momentum is heavily restricted such that {\it p}-wave and higher interactions are typically negligible.  Furthermore, the Pauli exclusion principle forbids identical fermions from interacting via {\it s}-wave collisions.  Thus, for an ultracold Bose-Fermi mixture (in which the fermions are identical), the
dominant interactions are {\it s}-wave boson-boson and
boson-fermion interactions.  It has been shown theoretically that a
repulsive Bose gas and a non-interacting Fermi gas co-existing in a radial waveguide can be coupled together by an
attractive boson-fermion interaction to form a self-trapped state
\cite{Karpiuk,BF_solitons}.   It is these `Bose-Fermi' solitons that we will consider in this paper.  We note
that the $^{87}$Rb-$^{40}$K system appears particularly well-suited
to support BF solitons since its natural boson-fermion interaction
is strongly attractive.  However, just as in the case of an attractively-interacting 3D Bose-Einstein condensate \cite{BEC_collapse,Gaussian_solitons,carr}, a 3D Bose-Fermi system is prone to collapse when the interspecies interaction
becomes too attractive \cite{Modugno2002,BF_collapse,roth,Karpiuk2005}.  While certain effects have been highlighted to raise the threshold for collapse, e.g. finite temperature effects in Bose-Fermi mixtures \cite{Liu2003} and Feshbach resonance management \cite{FRM}, possession of angular momentum \cite{ang_mom} and fragmentation \cite{frag} in Bose-Einstein condensates, the instability to collapse ultimately remains in the system.

By exploiting scattering resonances, the {\it p}-wave interaction
between fermions can now be experimentally engineered to significant values
\cite{P_wave_tuning}. It is thus the rationale of this work to
explore the way in which {\it p}-wave interactions may modify Bose-Fermi soliton
solutions.  Our framework to perform this is a variational approach
using a gaussian ansatz for the boson and fermion density distributions.  Via the Gross-Pitaveskii model for the bosons and the Thomas-Fermi approximation for the fermions, we derive an analytic form for the energy of this coupled system up to {\it p}-wave interactions (for the boson-fermion and fermion-fermion interaction) and from this we obtain the variational soliton solutions.  First we review the
properties of the soliton solutions in the absence of {\it p}-wave
interactions, in agreement with previous studies
\cite{Karpiuk,BF_solitons}, before
considering the modifications in the presence of {\it p}-wave
interactions.  Specifically we focus our results on
a $^{87}$Rb-$^{40}$K mixture, due to the naturally large and attractive {\it s}-wave scattering length between the species \cite{Rb-K,Modugno2002,Ferlaino2006},
and
fermion-fermion {\it p}-wave interactions, due to their capacity to be engineered experimentally \cite{P_wave_tuning}.

\section{Variational model of Bose-Fermi solitons}

\subsection{System overview and the variational ansatz}
We consider a degenerate gas of identical fermions co-existing with a
Bose-Einstein condensate of bosons, all at zero temperature.  Neglecting quantum and thermal fluctuations, we will model the fermion and boson gases within the mean-field picture.  Each gas is confined by an axially-homogeneous waveguide potential
$V_{\rm B \{ F \}}({\bf r})=\frac{1}{2}m_{\rm B\{F\}} \omega_{\rm
B\{F\}}^2 r^2$, where $\omega_{\rm B \{F\}}$ is the radial trap
frequency experienced by the bosons $\{$fermions$\}$ and $m_{{\rm B
\{F\}}}$ is the boson $\{$fermion$\}$ mass.  Due to the low energy of the atomic collisions, the {\it s}-wave and {\it p}-wave interactions are modelled by contact interactions characterised by a single length scale, the scattering length.   Within the Bose gas, the atoms interact predominantly via {\it s}-wave scattering with characteristic length $a_{\rm B}$ ({\it p}-wave interactions are negligible).  Within the Fermi gas, {\it s}-wave interactions are suppressed via the Pauli exclusion principle and the leading atomic interaction is {\it p}-wave with a  scattering length $a_{\rm F}$ \cite{roth}.  For overlapping clouds, the bosons and fermions additionally interact with each other, predominantly via the {\it s}-wave interaction, of lengthscale $a_{{\rm BF}s}$, but we will also include the corresponding {\it p}-wave interaction, with effective scattering length $a_{{\rm BF}p}$ \cite{roth}.

Bright solitons require an attractive interaction to enable
self-trapping of the wave.  Here we shall consider the case where
this interaction arises from the {\it s}-wave boson-fermion
coupling. A rudimentary requirement is thus that the boson and
fermion gases are overlapping in space and this enables us to assume the same ansatz for the boson and fermion density distributions.  We will assume that the radial profile of the fermion and boson gases is a gaussian.  This is an exact result in the quasi-1D limit (formally expressed as $\hbar
\omega_{\rm B} \gg \mu_{\rm B}$ and $\hbar \omega_{\rm F} \gg \mu_{\rm
F}$, where $\mu_{\rm B}$ and $\mu_{\rm F}$ are the chemical
potentials of the boson gas and fermion gas, respectively \cite{Das2003}) for which the radial profile coincides with the gaussian ground harmonic oscillator state.  In the following, we do not impose a fixed size to our gaussian profile but simply require the transverse profile to approximate a gaussian shape.  Thus our approach can apply to systems that lie outside of this dimensionality condition.  However, as is shown later, the radial sizes of the solitons remain close to the radial harmonic oscillator length $l_{\rm ho}$ throughout.

The most obvious choice for a suitable axial profile is, by analogy to 1D bright bosonic soliton result, a sech-profile \cite{Pethick}.  However, with this choice we are
unable to obtain analytic solutions for the variational energies.  Karpiuk {\it et al.} \cite{Karpiuk} pursued this choice numerically.  Instead, to obtain an analytic form for the variational energies, we employ a gaussian axial profile. From studies of bright
BEC solitons it has been shown, firstly, that sech and gaussian axial profiles give very similar results, and secondly, that both forms of ansatz give very good agreement with more precise theoretical treatments, e.g. numerical solutions of the Gross-Pitaevskii equation \cite{Gaussian_solitons,carr}.  We will thus consider the boson
$\{$fermion$\}$ density $n_{\rm B\{F\}}$ to have a
cylindrically-symmetric gaussian profile,
\begin{equation}
n_{\rm B\{F\}}({\bf r})=\frac{N_{\rm B\{F\}}}{\pi^{3/2} L_z
L_r^2}\exp\left(-\frac{z^2}{L_z^2}\right)\exp
\left(-\frac{r^2}{L_r^2}\right), \label{eqn:var_density}
\end{equation}
where $L_r$ and $L_z$ are the radial and axial sizes, respectively,
and $N_{\rm B}$$_{\{{\rm F}\}}$ is the numbers of bosons
$\{$fermions$\}$. We consider that $L_r$ and $L_z$ are common for both the bosons and the fermions.

Note that the validity of the mean-field Gross-Pitaevskii model for the boson gas requires that $N_{\rm B}\gg 1$.  Furthermore, our description of the Fermi gas component is based on the Thomas-Fermi approximation which is valid for $N_{\rm F}^{1/3}\gg1$ \cite{Pethick}.  Hence our approach is strictly valid only in the regime of large $N$ and we will only consider parameters that satisfy this.

\subsection{Energetics of the system}
We will consider the total energy density of the Bose-Fermi state
$\varepsilon[n_{\rm B}, n_{\rm F}]$ to be the sum of the boson
contribution $\varepsilon_{\rm B}[n_{\rm B}]$, fermion contribution
$\varepsilon_{\rm F}[n_{\rm F}]$ and boson-fermion term
$\varepsilon_{\rm BF}[n_{\rm B}, n_{\rm F}]$ \cite{roth}.  We will proceed by modelling each energy contribution of the gaussian wavepackets in turn.  Note that the
energy is the volume integral of the corresponding energy density
$E=\int \varepsilon[n] dV$.
For a different choice of ansatz and in the absence of {\it p}-wave interactions, Karpiuk {\it et al.} \cite{Karpiuk} followed a similar variational approach to explore Bose-Fermi soliton solutions.

\subsubsection{Bosonic energy contribution}
The energy density of the zero-temperature boson gas, interacting via {\it s}-wave interactions of scattering length $a_{\rm B}$, is provided by the Gross-Pitaevksii model
\cite{Pethick},
\begin{equation}
\varepsilon_{\rm B}[n_{\rm B}]=\frac{\hbar^2}{2m}|\nabla
\sqrt{n_{\rm B}}|^2+V_{\rm B}({\bf r}) n_{\rm B} +\frac{2\pi \hbar^2
a_{\rm B}}{m_{\rm B}}n_{\rm B}^2. \label{eqn:GP_energy_functional}
\end{equation}
The terms of the right-hand side represent, respectively, the kinetic, potential and interaction energies of the boson gas.
For convenience we will express length in
terms of the boson harmonic oscillator length $ l_{\rm
ho}=\sqrt{\hbar/m_{\rm B} \omega_{B}}$ and adopt dimensionless
variational lengthscales $l_z=L_z/l_{\rm ho}$ and $l_r=L_r/l_{\rm
ho}$. Furthermore, we rescale energy by the bosonic harmonic
oscillator energy $\hbar \omega_{\rm B}$ via $E\rightarrow E/(\hbar
\omega_{\rm B})$. Substituting $n_{\rm B}({\bf r})$ into
Eq.~(\ref{eqn:GP_energy_functional}) and integrating over space, one
arrives at the well-known expression for the total boson energy \cite{Pethick},
\begin{equation}
\frac{E_{\rm B}}{N_{\rm B}}=\frac{1}{2}\left(\frac{1}{l_r^2}+\frac{1}{2l_z^2}
\right)+\frac{1}{2}l_r^2+\frac{1}{\sqrt{2\pi}} \frac{a_{\rm
B}}{l_{\rm ho}}\frac{N_{\rm B}}{l_z l_r^2}.
\label{eqn:boson_energy}
\end{equation}
From this the presence of the collapse instability for
attractively interacting BECs ($a_B<0$) is clear: the interaction term (negative) dominates over all other (positive) energy
contributions as the size of the wavepacket tends towards zero.  While meta-stable states can be formed for certain parameters, the lowest energy state described by this equation is always a collapsed state of zero width.

\subsubsection{Fermionic energy contribution}
Determination of the mean-field fermionic energy density in general
requires solving $N_{\rm F}$ coupled Hartree-Fock equations. In this
manner, Karpiuk {\it et al.} \cite{Karpiuk} successfully modelled BF
soliton solutions, but the approach is only tractable for of the
order of $10$ fermions.  Instead, we will adopt an analytic form
for the fermion energy density derived by Roth and Feldmeier \cite{roth}.
Employing the Thomas-Fermi approximation, this approach follows from
deriving the energy density of a homogeneous fermionic system and
replacing the fixed density with $n_{\rm F}({\bf r})$ (thus
neglecting the contribution to the energy from density gradients).  The Thomas-Fermi approximation is valid for the equlibrium state of a fermi gas when the fermion wavelength is much smaller than the system size.  For a trapped fermi gas, it can be shown that this is satisfied when $N_{\rm F}^{1/3}\gg 1$ \cite{Pethick}.  In agreement with this, the Thomas-Fermi approximation has been shown to give an excellent mean-field description of BF mixtures
(in close agreement with Hartree-Fock calculations) for $N_{\rm
F}\sim1000$ \cite{nygaard,roth}.  Via this
approach, the energy density of the fermions (potential, kinetic and
{\it p}-wave interaction terms, respectively) is \cite{roth},
\begin{equation}
\varepsilon_{\rm F}[n_{\rm F}]=V_{\rm F}({\bf r})n_{\rm F}+\frac{3
\hbar^2(6 \pi^2)^{2/3}}{10m_{\rm F}}n_{\rm F}^{5/3}+\frac{(6
\pi^2)^{5/3}}{5\pi m_{\rm F}}\hbar^2 a_{\rm F}^3 n_{\rm F}^{8/3},
\label{eqn:fermion_energy_density}
\end{equation}
where $a_{\rm F}$ is the  {\it p}-wave contact interaction
between fermions \cite{roth}. Inserting the gaussian profile
$n_{\rm F}({\bf r})$ and integrating gives the total fermionic
energy,
\begin{eqnarray}\label{eqn:fermion_energy}
\frac{E_{\rm F}}{N_{\rm F}}&=&\alpha \frac{m_{\rm B}}{m_{\rm
F}}\left(\frac{N_{\rm F}}{l_z l_r^2}\right)^{2/3}+\frac{1}{2}\frac{m_{\rm F}}{m_{\rm B}}\left(\frac{\omega_{\rm
F}}{\omega_{\rm B}}\right)^2 l_r^2\\&+&4\beta \frac{m_{\rm
B}}{m_{\rm F}}\left(\frac{a_{\rm F}}{l_{\rm
ho}}\right)^3\left(\frac{N_{\rm F}}{l_z l_r^2}\right)^{5/3}
\nonumber
\end{eqnarray}
where $\alpha=(9/50)6^{2/3}(3/5)^{1/2}\pi^{1/3}\approx0.6743$ and
$\beta=(3/160\pi^4)(3\pi/8)^{1/2}(6\pi^2)^{5/3}\approx0.1880$.

\subsubsection{Bose-Fermi energy contribution}
Again under the Thomas-Fermi approximation, the interaction energy
density between the bosons and fermions $\varepsilon_{\rm BF}[n_{\rm
B}, n_{\rm F}]$ is \cite{roth},
\begin{equation}
\varepsilon_{\rm BF}[n_{\rm B}, n_{\rm F}]=\frac{2\pi \hbar^2 a_{{\rm
BF}s}}{\mu} n_{\rm B} n_{\rm F}+\frac{(6 \pi^2)^{5/3}}{20 \pi
\mu}\hbar^2 a_{{\rm BF}p}^3 n_{\rm B} n_{\rm F}^{5/3},
\label{eqn:interaction_energy_density}
\end{equation}
where $a_{\rm BF {\it s}}$ is the {\it s}-wave boson-fermion scattering
coefficient, $a_{\rm BF{\it p}}$ is the {\it p}-wave scattering
length and $\mu=m_{\rm B} m_{\rm F}/(m_{\rm B}+m_{\rm F})$.  For the density profiles (\ref{eqn:var_density}) the boson-fermion interaction energy is,
\begin{equation}
\frac{E_{\rm BF}}{N_{\rm B}}=\frac{1}{\sqrt{2\pi}}\frac{a_{\rm BF{\it s}}}{l_{\rm
ho}}\frac{m_{\rm B}}{\mu} \frac{N_{\rm F}}{l_z
l_r^2}+\beta\frac{m_{\rm B}}{\mu} \left(\frac{a_{\rm
BF{\it p}}}{l_{\rm ho}}\right)^3 \left(\frac{N_{\rm F}}{l_z
l_r^2}\right)^{5/3} \label{eqn:interaction_energy}.
\end{equation}

\subsection{Energy landscapes and obtaining the variational solutions}
The total energy of our gaussian Bose-Fermi wavepackets is given by $E=E_{\rm B}+E_{\rm F}+E_{\rm BF}$.  Upon fixing the experimental parameters (atom masses, atom numbers, scattering
lengths and trap frequencies), the energy becomes confined to being a function of only the wavepacket size parameters $l_r$ and $l_z$.  This function $E(l_r,l_z)$ can be visualised as an energy landscape in which the presence of an energy minimum represents a variational solution.  In practice, we numerically define this energy landscape and perform a simple computational search for such energy minima.  Note that an unphysical solution can occur at the origin representing the effect of collapse.  It is unphysical in the sense that a real Bose-Fermi system cannot shrink to zero size; in reality, a collapse will eventually become halted by the surge of three-body losses as the gas densities rise.  At most, only one physical solution is ever present in these energy landscapes.  We denote the coordinates of such a variational solution by the coordinates $l^0_z$ and $l^0_r$.

Where we map out regions of soliton solutions within a particular parameter space, e.g., $a_{{\rm BF}s}-N_{\rm F}$ space, this is done by randomly sampling combinations of these parameters.  We typically restrict our numerical search to landscapes of extent $[0,2]~l_r \times [0,10]~l_z$.

We have verified that our methodology gives very close agreement with the variational stability diagram presented in Fig.~7 of \cite{Karpiuk}.  Our results deviate by less than $10\%$, a deviation which is to be expected given the different ansatz employed.

\subsection{Analytical limit: no {\it p}-wave interactions and $N_{\rm B}\gg N_{\rm F}$}

We can gain a simplified analytic form for the total variational energy if we neglect {\it p}-wave interactions ($a_{\rm F}=a_{{\rm BF}p}=0$) and assume $N_{\rm B}
\gg N_{\rm F}$.  The latter condition renders the fermion-fermion energy terms negligible and makes significant only terms involving $N_{\rm B}$.  The total variational energy then reduces to,
\begin{equation}
\frac{E}{N_{\rm B}}=\frac{1}{2l_r^2}+\frac{1}{4l_z^2}
+\frac{l_r^2}{2}+\left(N_{\rm B} \frac{a_{\rm B}}{l_{\rm
ho}}+\frac{a_{{\rm BF}s}}{l_{\rm ho}} \frac{m_{\rm B}}{\mu}N_{\rm
F}\right)\frac{1}{\sqrt{2\pi} l_z l_r^2}.
\end{equation}
This form will enable us to gain physical intuition of the system and provide simple criteria for the existence of stable Bose-Fermi solitons.  Importantly, it has the same form as the gaussian variational energy of a purely bosonic
gas \cite{Gaussian_solitons}, but with an effective {\it s}-wave
scattering length given by \cite{Karpiuk2005},
\begin{equation}
a_{\rm eff}=a_{\rm B}+a_{{\rm BF}s}\frac{N_{\rm F}}{N_{\rm
B}}\left(\frac{m_{\rm F}+m_{\rm B}}{m_{\rm F}}\right).
\label{eqn:net_scat_length}
\end{equation}
A rudimentary requirement for the ability to self-trap is that the
net interactions are attractive ($a_{\rm eff}<0$).  This places a lower bound on the ratio $N_{\rm F}/N_{\rm B}$ for which Bose-Fermi solitons can be self-supported,
\begin{equation}
\frac{N_{\rm F}}{N_{\rm B}}>-\frac{a_{\rm B}}{a_{{\rm
BF}s}}\left(\frac{m_{\rm F}}{m_{\rm F}+m_{\rm B}}\right).
\label{eqn:dispersion_criteria}
\end{equation}

Furthermore, bright bosonic solitons are established to collapse when the scattering length is less than the critical value $a^c_{\rm B}=k_c l_{\rm ho}/N_{\rm
B}$.  For a 3D gaussian wavepacket the dimensionless coefficient has been shown to be $k_c=-0.778$
\cite{Gaussian_solitons} (for other shapes of wavepacket the value differs but remains of the order of unity).  This leads to an upper bound for the ratio $N_{\rm F}/N_{\rm B}$ in order to prevent collapse of the system,
\begin{eqnarray}
\frac{N_{\rm F}}{N_{\rm B}}<\frac{k_c l_{\rm ho}}{N_{\rm B} a_{{\rm
BF}s}}\left(\frac{m_{\rm F}}{m_{\rm F}+m_{\rm B}} \right)-\frac{a_{\rm
B}}{a_{{\rm BF}s}}\left(\frac{m_{\rm F}}{m_{\rm F}+m_{\rm B}}\right).
\label{eqn:collapse_criteria}
\end{eqnarray}

From Eqs.~(\ref{eqn:dispersion_criteria},\ref{eqn:collapse_criteria}) it is evident that the soliton solutions exist within a ``window" of fermion atom number $N_{\rm F}$ whose width is,
\begin{equation}
\Delta N_{\rm F}=\frac{k_c  l_{\rm ho}}{a_{{\rm BF}s}}\left( \frac{m_{\rm F}}{m_{\rm F}+m_{\rm B}} \right).
\label{eqn:width}
\end{equation}

Equations ~(\ref{eqn:dispersion_criteria}), (\ref{eqn:collapse_criteria}) and (\ref{eqn:width}) provide us with an estimate for the locality and range over which BF solitons solutions may exist.
Equation (\ref{eqn:width}) indicates that the width of the soliton bands can be extended by employing weaker radial trapping and weaker Bose-Fermi scattering length.  However, we will find that the soliton bands are even more restricted if the condition $N_{\rm B}/N_{\rm F}\gg1$ is removed.  Indeed, the above equations predict the existence of soliton solutions at the native Bose-Fermi scattering length $a_{{\rm BF}s}=-215a_0$.  However, as we shall see, the full variational results predict that the soliton bands become vanishing narrow at this scattering length.

\section{Results}
While our analytical results presented so far are applicable to any Bose-Fermi species, we will focus our ensuing results on a $^{87}$Rb and $^{40}$K mixture   due to its naturally strong and attractive Bose-Fermi coupling and its prominent experimental occurence to date \cite{Rb-K,Modugno2002}.  We will assume that the atoms are spin-polarized and confined (in the radial direction) by a magnetic trap such that the trap frequencies are related via $\omega_{\rm F}/\omega_{\rm B}=(m_{\rm B}/m_{\rm F})^{1/2}$.  Throughout our results we fix the boson-boson {\it s}-wave scattering length to be the experimentally measured value of $a_{\rm B}=99a_0$ \cite{Kempen2002}, where $a_0=5.3\times10^{-11}$m is the Bohr radius.  We will consider the radial trap frequency, boson and fermion atom numbers, and remaining scattering lengths to be variables.  Note that scattering lengths can typically be experimentally varied over many orders of magnitude by using magnetic and optical fields to access inter-atomic scattering resonances.  In the cases where we fix the boson-fermion {\it s}-wave scattering length, we take it to be its native value of $a_{{\rm BF}s}=-215a_0$, as measured by Ferlaino {\it et al.}  \cite{Ferlaino2006}.

First we will explore the soliton solutions in the absence of {\it p}-wave interactions.  Following this we will consider how {\it p}-wave interactions between fermions modify the soliton solutions.  We will not explicitly present results for finite boson-fermion {\it p}-wave interactions but will comment on how they affect the system in Section IV.

\subsection{Absence of {\it p}-wave interactions}

\begin{figure}[t]
\includegraphics[width=7cm,clip=true]{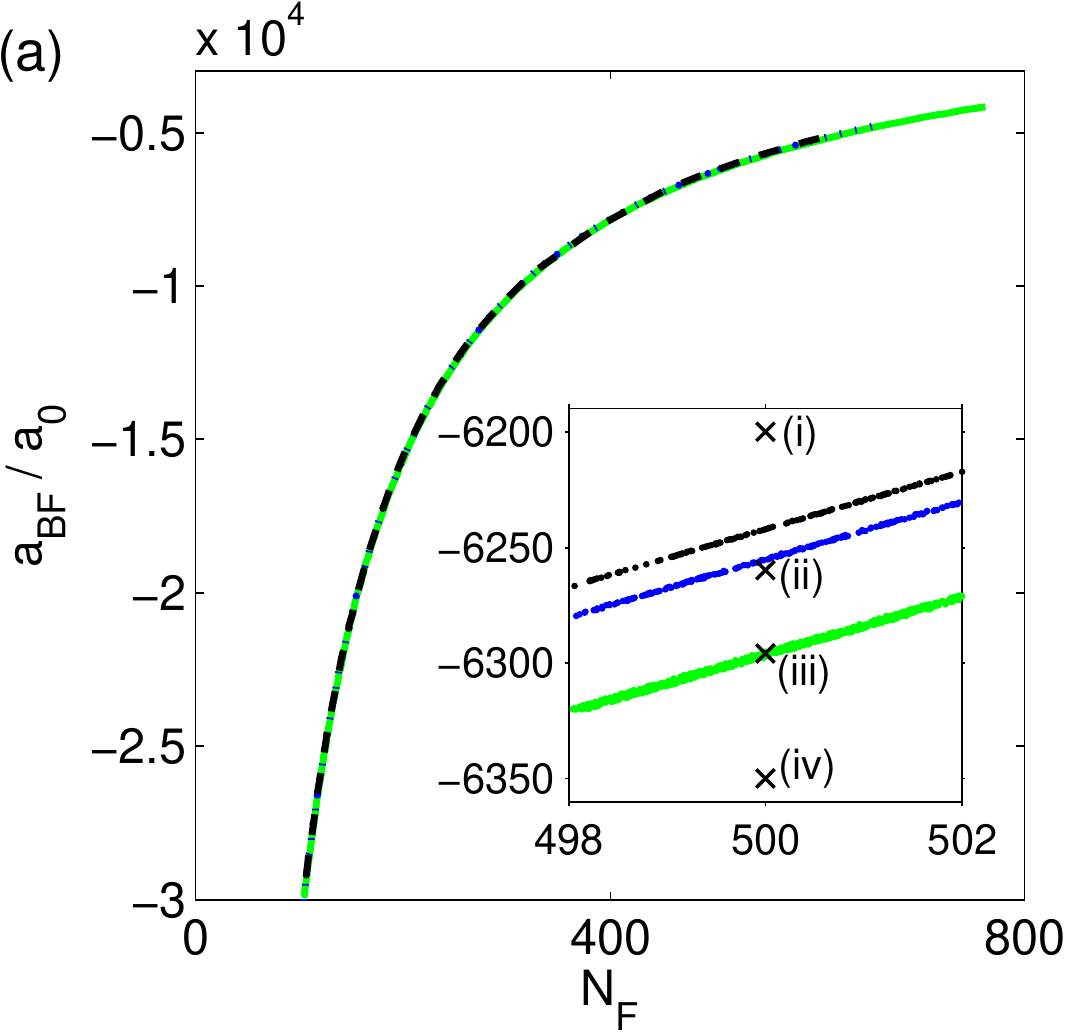}\\
\includegraphics[width=6cm, clip=true]{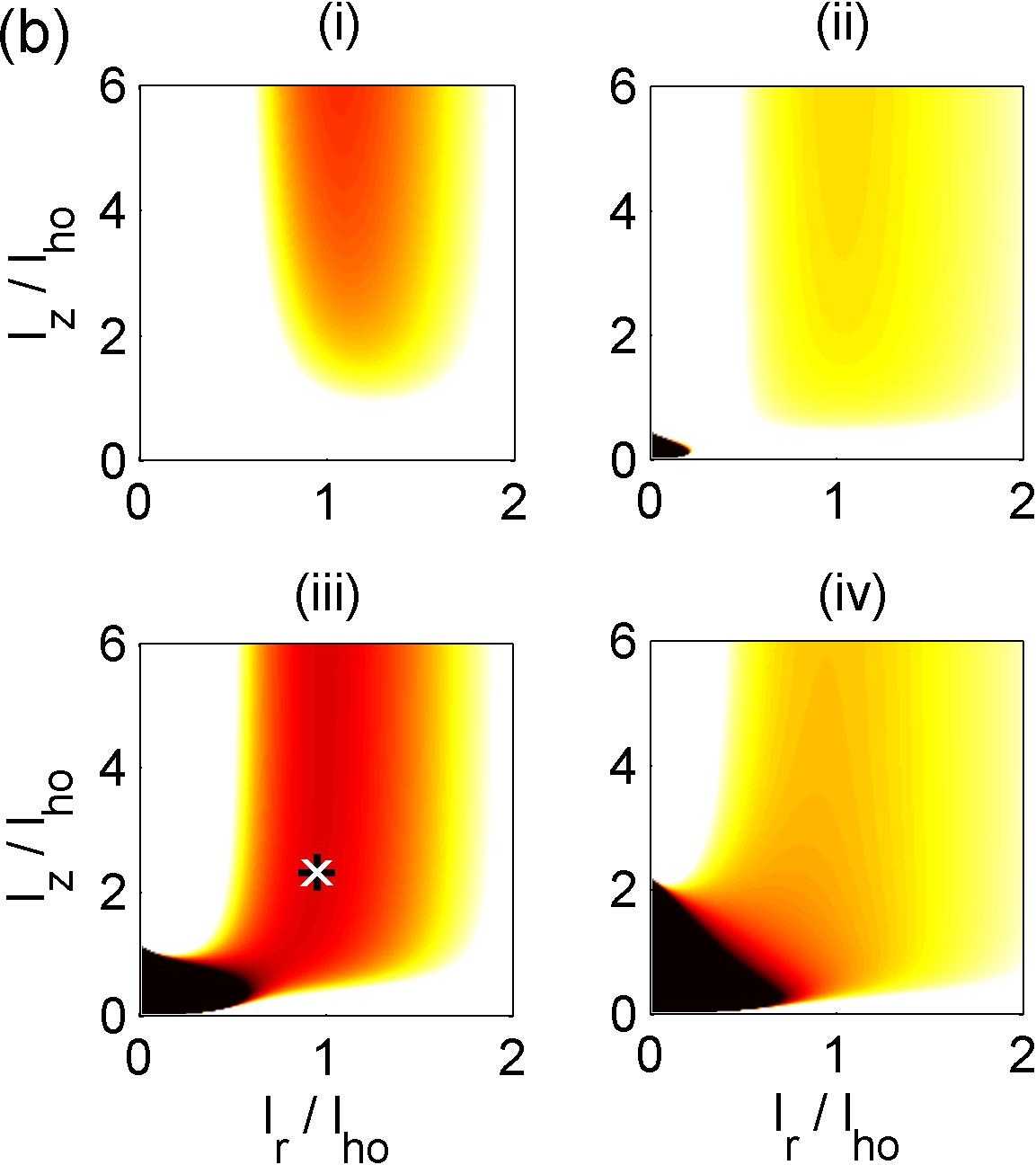}
\caption{(a) The bands of soliton solutions in $a_{{\rm BF}s}-N_{\rm F}$ space for a ${^{87}}$Rb-${^{40}}$K mixture with zero {\it p}-wave interaction. We present a fixed number of bosons $N_{\rm B}=10^5$ and various trap frequencies
$\omega_{\rm B}/2\pi$ =$10$ (green solid), $100$ (blue dotted),
$1000$ (black dashed) Hz.  The inset provides a close up of the soliton bands. (b) Energy landscapes ($\omega_{\rm
B}/2\pi=10$Hz, $N_{\rm F}=500$ and $N_{\rm B}=10^5$) showing four distinct regimes: (i) well above the
soliton band ($a_{{\rm BF}s}=-6200a_0$), (ii) just above the band
($a_{{\rm BF}s}=-6260a_0$), (iii) within the band ($a_{{\rm
BF}s}=-6296a_0$) and (iv) below the band. ($a_{{\rm BF}s}=-6350a_0$). These plots corresponds to the crosses in (a). In
(iii) the soliton solution is highlighted by the white cross.}
\label{fig:vary_omega}
\end{figure}

\subsubsection{Soliton bands in $N_{\rm F}-a_{{\rm BF}s}$ space}
Figure \ref{fig:vary_omega}(a) presents the soliton solutions in the
parameter space of $a_{{\rm BF}s}-N_{\rm F}$ for $10^5$ bosons
and various radial trap frequencies [$\omega_{\rm B}/2\pi=10$ (green solid region), $100$ (blue dotted region) and $1000$ Hz (black dashed region)]. For each trap
frequency, the soliton solutions exist in a narrow band in the
$a_{{\rm BF}s}<0$ half-plane. (Note that on the scale of the figure these bands appear as lines).  Above each band, the system is unstable to dispersion and below the band it is unstable to collapse.  Each band scales approximately as
$-1/N_{\rm F}$, as indicated by rearranging Eqs.
(\ref{eqn:dispersion_criteria}) and (\ref{eqn:collapse_criteria}). The
bands are weakly dependent on trap frequency.  Indeed their differences are only apparent on a much more magnified scale (inset of Fig.\ref{fig:vary_omega}(a)).  As $\omega_{\rm B}$ is
increased, the bands shift upwards in $a_{{\rm BF}s}$ and become
slightly narrower.  The latter change is in qualitative agreement with
equations
(\ref{eqn:dispersion_criteria}-\ref{eqn:width}) which
predict the band width to scale in proportion to $l_{\rm
ho}=\sqrt{\hbar/m\omega_{\rm B}}$.
Although not visible in Figure \ref{fig:vary_omega}(a), the bands
become progressively narrower as $N_{\rm F}$ increases. Indeed,
beyond some critical fermion number $N^{\rm crit}_{\rm F}$ we cannot
detect further solutions.  This marks an important difference between the approximated analytic predictions of Eqs.~(\ref{eqn:dispersion_criteria})-(\ref{eqn:width}) and the full variational solutions.  For example, for trap frequencies
$\omega_{\rm B}/2\pi=10, 100$ and $1000$, $N^{\rm crit}_{\rm F}
\approx 760, 660$ and $620$, respectively.

Within the soliton bands, the soliton radial size $l_r^0$ remains
close to $l_{\rm ho}$ throughout. The axial size $l_z^0$ is
infinite at its dispersive boundary and reduces as $a_{{\rm BF}s}$ is
made more attractive, with the soliton becoming almost spherical at
the point of collapse.  This is qualitatively similar to the case
for BEC bright solitons \cite{Gaussian_solitons}.

\begin{figure}[t]
\includegraphics[width=\columnwidth,clip=true]{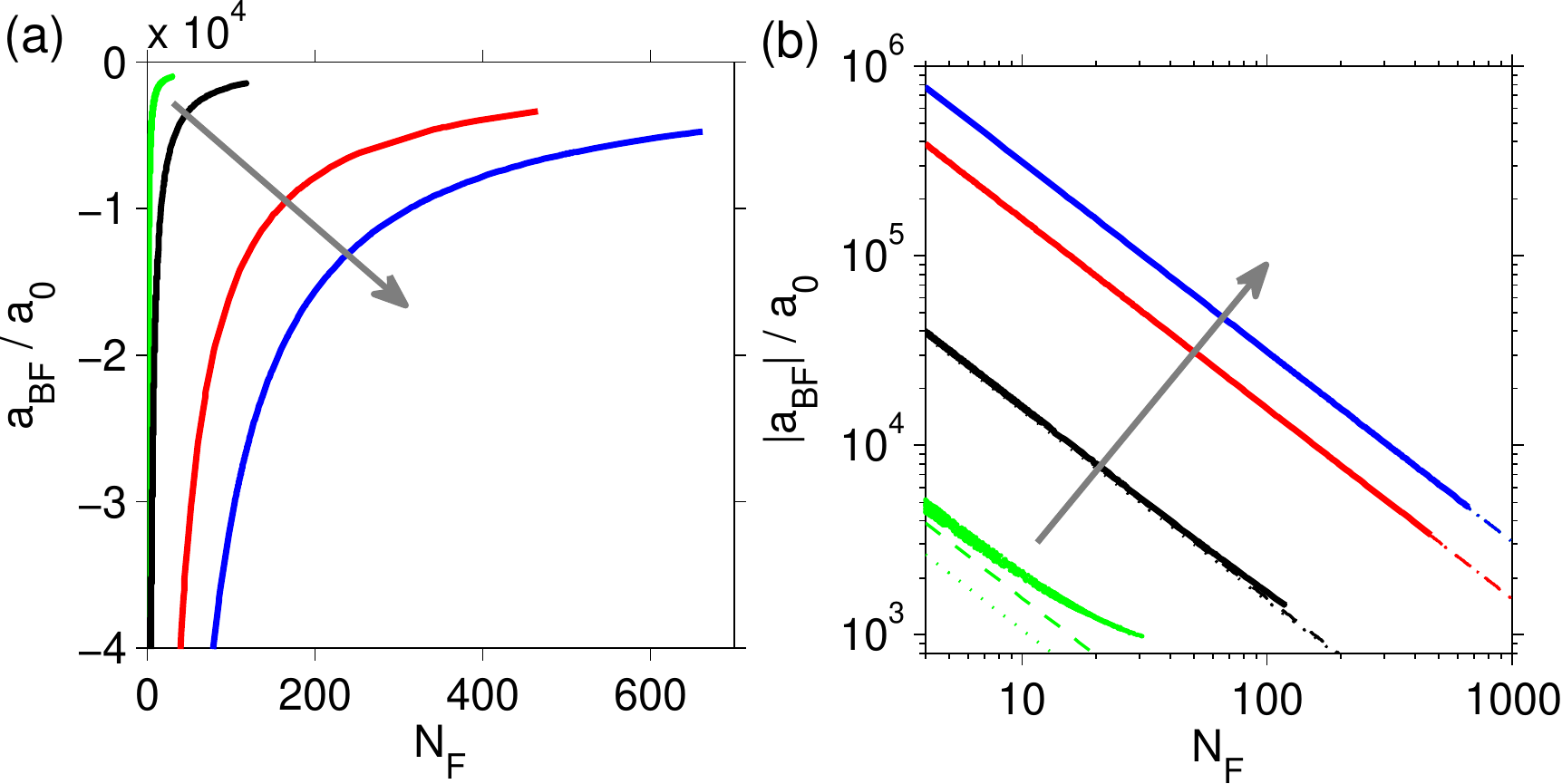}
\caption{ (a) The bands of soliton solutions in $a_{{\rm BF}s}-N_{\rm F}$ space for no {\it p}-wave contributions.  We consider fixed
trapping $\omega_{\rm B}/2\pi=100$Hz and various boson numbers of
$N_{\rm B}=500$ (green), $5,000$ (black), $50,000$ (red) and
$100,000$ (blue). (b) Log-log plot of the soliton bands in (a), with
the analytic predictions of Eqs.~(\ref{eqn:dispersion_criteria})
(dashed line) and (\ref{eqn:collapse_criteria}) (dotted line).  The grey arrows indicate the direction of increasing $N_{\rm B}.$}
\label{fig:vary_NB}
\end{figure}

Figure \ref{fig:vary_NB} shows how the soliton bands change for different number of bosons (in both linear (a) and logarithmic plots (b)).  For increasing $N_{\rm B}$ the bands shift to more negative $a_{{\rm BF}s}$ and larger $N_{\rm F}$, and the bands becomes wider as $N_{\rm B}$ is reduced.

Our numerical results predict that, for the numbers of bosons and fermions permitted by our model ($\gg 1$), soliton solutions do not occur at the native Bose-Fermi scattering length $a_{{\rm BF}s}=-215a_0$.  The soliton solutions exist only for scattering lengths $a_{{\rm BF}s}\ll-215a_0$.  However, our results show that as $N_{\rm B}$, and by association $N_{\rm F}$ (since the ratio $N_{\rm B}/N_{\rm F}$ must remain within narrow bounds for soliton-supporting conditions to be met, see, e.g., Eqs.~(\ref{eqn:dispersion_criteria})-(\ref{eqn:width})) is decreased the soliton solutions extend to smaller $|a_{{\rm BF}s}|$.  Indeed, if one were to extrapolate our predictions (beyond its strict regime of validity) to lower atom number, one could imagine that the soliton bands would reach $a_{{\rm BF}s}=-215a_0$.   Indeed, using a model valid for low atom numbers, Karpiuk {\it et al.} \cite{Karpiuk} predict the existence of BF solitons at the native $a_{{\rm BF}s}=-215a_0$ for very low atom numbers.

\subsubsection{Energy landscape regimes}
To gain physical insight into the system, in Fig.~\ref{fig:vary_omega}(b)(i)-(iv) we present energy landscapes of four distinct regimes in this parameter space.  While we present the landscapes for a specific set of parameters ($\omega_{\rm B}/2\pi=10$ Hz, $N_{\rm F}=500$ and $N_{\rm B}=10^5$) the qualitative behaviour is generic.  The location of each case is indicated in the inset of Fig.~\ref{fig:vary_omega}(a) by crosses.  These four regimes [Fig.~\ref{fig:vary_omega}(i)-(iv)] are as follows:
\begin{list}{\labelitemi}{\leftmargin=1em}
\item (i) Sufficiently above the soliton band, the net contact
interactions are repulsive (positive) and the energy landscape
consists of a downward `chute' aligned along the $l_z$ axis. Any
wavepacket subjected to this system will disperse axially.

\item(ii) Just above the soliton band the net interactions become attractive
(negative) and compete with the positive energy terms.  The
chute remains but the global energy minimum is now at the
origin, where the energy diverges to $-\infty$. The attraction is
insufficiently strong to support a soliton.

\item(iii) Within the band, the play-off between the interspecies attraction
and repulsive zero-point kinetic energy leads to a local energy
minimum at [$l^0_r, l^0_z$]
 (highlighted by the white cross), corresponding to the self-trapped
soliton solution.
\item(iv) Below the soliton band the attractive interactions dominate the kinetic energy such that the only energy minimum is
the global minimum at the origin, representing the collapse
instability of the system.
\end{list}

\subsubsection{Comparison to the simplified limit of Eqs.~(\ref{eqn:dispersion_criteria}) and (\ref{eqn:collapse_criteria})}
In Fig.~\ref{fig:vary_NB}(b) we present a comparison between the variational predictions for the soliton bands and the simplified analytic estimates provided by
Eqs.~(\ref{eqn:dispersion_criteria}) (dashed line) and
(\ref{eqn:collapse_criteria}) (dotted line).  For ease of observing the differences between the two methods, the data is presented on a log-log plot.  For low number of bosons ($N_B=500$) the variational
solutions deviate from the predictions. However, for larger boson
number the agreement is excellent. This is to be expected since
these predictions assume $N_{\rm B}/N_{\rm F} \gg 1$.  Indeed for
$N_B=5,000$ and above, the simplified forms give very good
predictions for the regimes of soliton solutions
(indistinguishable from the full variational results on the scale of this
figure).

An important difference, however, is that according to the analytical result, the width of the soliton band decreases as $1/N_{\rm F}$, becoming vanishingly small only as $N_{\rm F} \rightarrow \infty$.  In contrast, the full variational solutions disappear beyond a finite $N_{\rm F}$.  Furthermore, numerically, the bands increase in width as $N_{\rm B}$ is increased, whereas the analytical prediction for the band width is independent of $N_{\rm B}.$

\subsection{The role of {\it p}-wave fermion interactions}
\begin{figure}[t]
\centering
\includegraphics[width=6.5cm,clip=true]{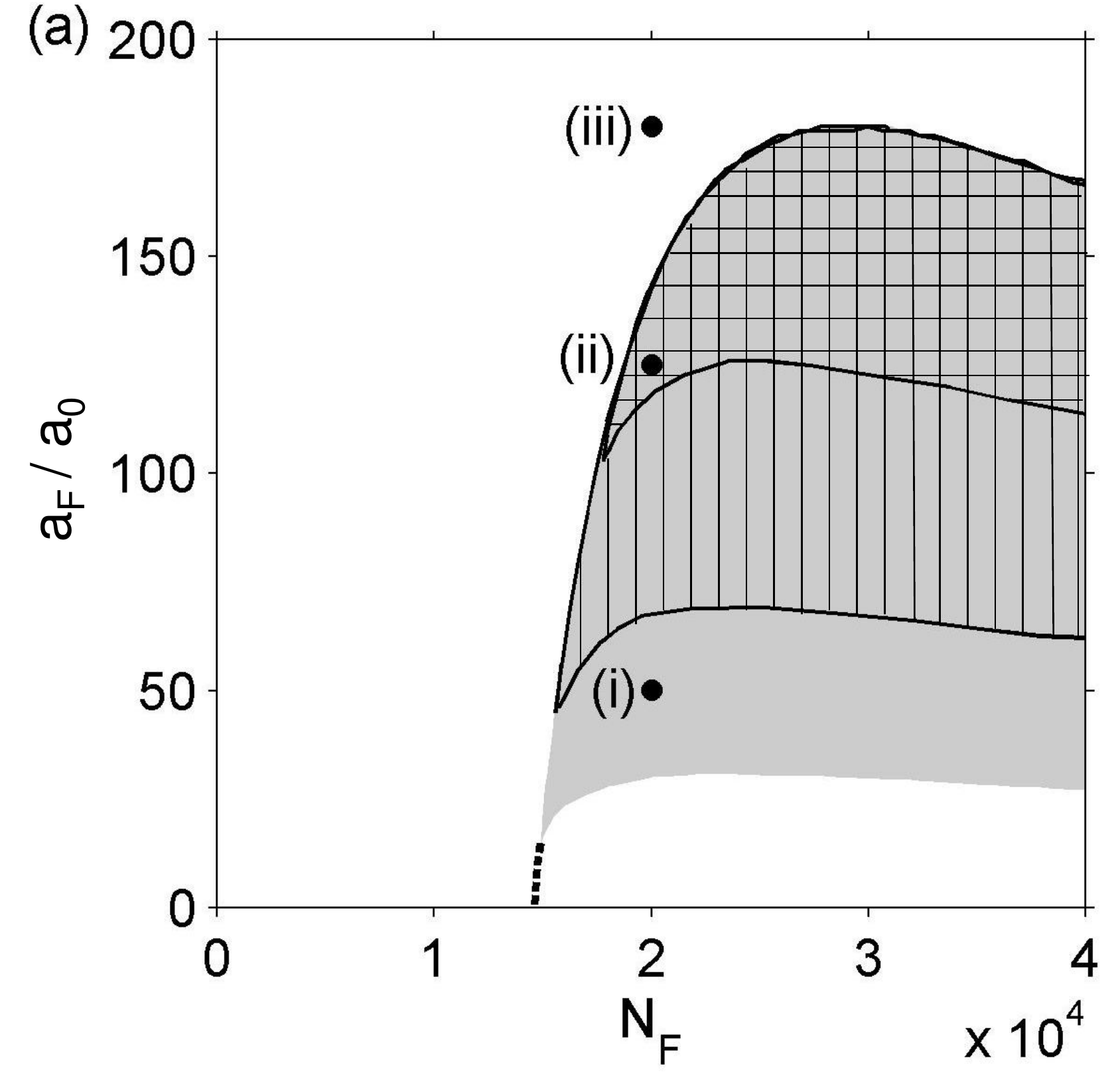}\\
\includegraphics[width=8cm,clip=true]{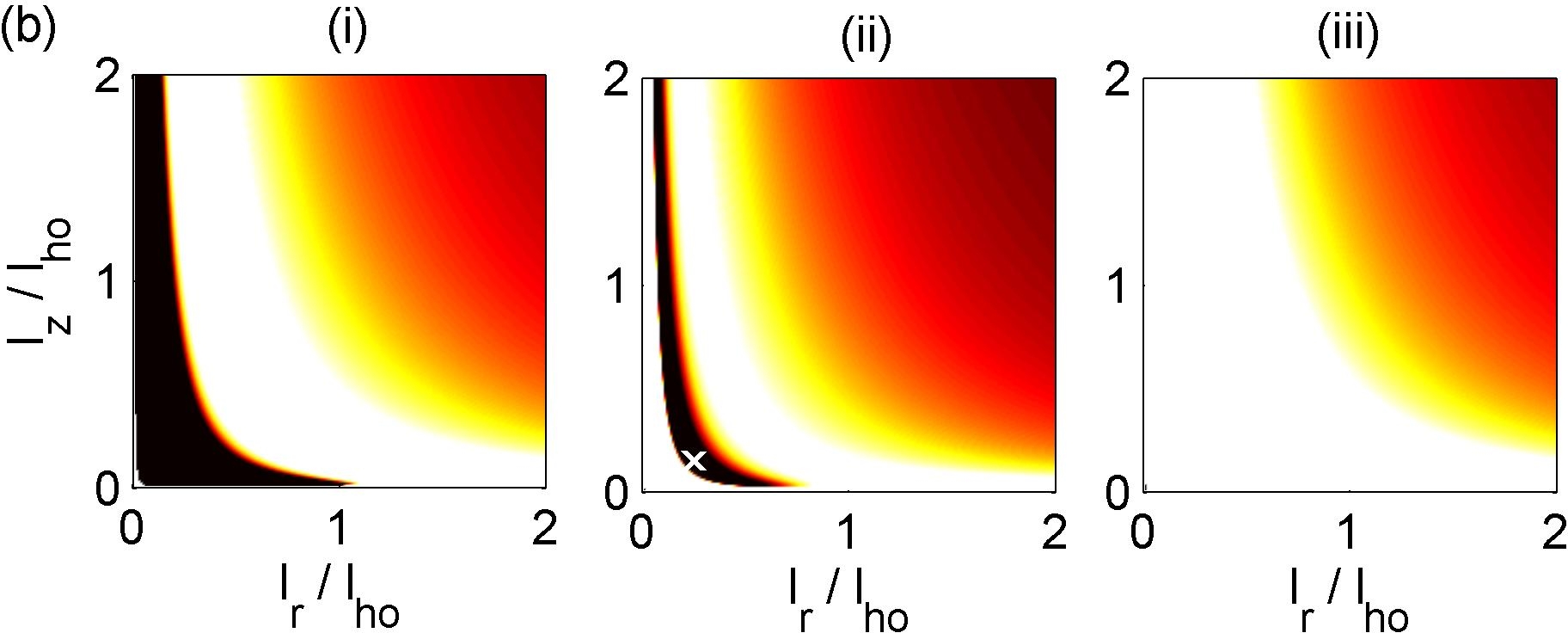}\\
\caption{(a) Soliton bands in $a_{\rm F}-N_F$ space for
boson-fermion interaction $a_{{\rm BF}s}=-215a_0$ in presence of fermion-fermion {\it p}-wave interactions. The number of bosons
is fixed to $N_{\rm B}=100,000$ and we present trap frequencies of
$\omega_{\rm B}/2\pi=10$ (horizontal thatch), $100$ (vertical
thatch) and $1,000$ Hz (grey region). (b) Energy landscapes
(for $\omega_{\rm B}/2\pi=100$ and $N_{\rm F}=20,000$) of (i)
below the band ($a_{\rm F}=50a_0$) (ii) within the band ($a_{\rm
F}=125a_0$) and (iii) above the band ($a_{\rm F}=180a_0$).  The
locations of these cases are indicated in plot (a).}
\label{fig:aF_vs_NF_vary_wB}
\end{figure}
\subsubsection{Soliton bands in $a_{\rm F}-N_{\rm F}$ space}
We now consider the presence of fermion-fermion ({\it p}-wave)
interactions.  We fix the boson-fermion {\it s}-wave interaction to its
natural value $a_{{\rm BF}s}=-215a_0$ and explore the parameter space
of $a_{\rm F}-N_{\rm F}$.  The results are shown in
Fig.~\ref{fig:aF_vs_NF_vary_wB} for fixed boson number $N_{\rm B}=10^5$ and various trap frequencies [$\omega_{\rm B}/2\pi=10$ (horizontal thatch), $100$ (vertical
thatch) and $1000$ Hz (grey region)].  Recall that in the absence of {\it p}-wave
interactions, no solitons were obtained for $a_{{\rm BF}s}=-215a_0$. In
contrast, in the presence of repulsive {\it p}-wave fermion-fermion interactions, we now see extensive regions of soliton solutions.  The regions become larger for increased trap frequency.  This change in size occurs due to a shift in the lower boundary of the regions; the upper boundary is insensitive to $\omega_{\rm B}$, as can be seen in Fig.~\ref{fig:aF_vs_NF_vary_wB}(a).

\begin{figure}[t]
\includegraphics[width=6cm,clip=true]{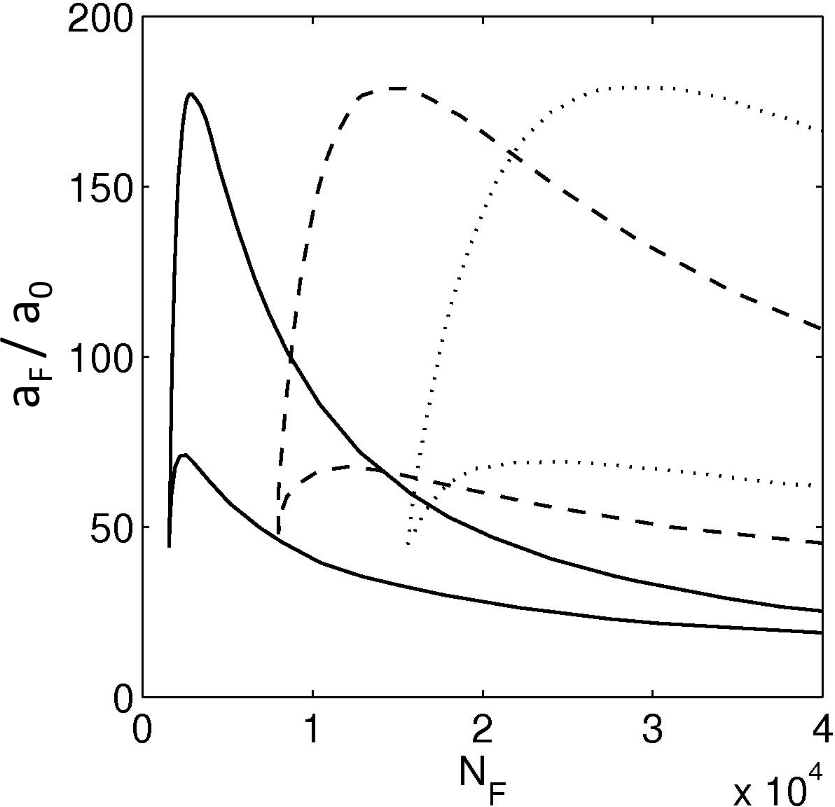}\\
\caption{Soliton bands in $a_{\rm F}-N_F$ space for fixed trap
frequency $\omega_{\rm B}/2\pi=100$ Hz and boson numbers of $N_{\rm
B}=10,000$ (solid), $50,000$ (dashed) and $100,000$ (dotted). }
\label{fig:aF_vs_NF_vary_NB}
\end{figure}

This enhanced stability when {\it p}-wave fermion-fermion interactions are included
arises from the fact that the fermion-fermion interaction term in
Eq. (\ref{eqn:fermion_energy}) scales as $l^{-5}$. Thus, if $a_{\rm
F}>0$, this term ensures that the energy diverges to positive values
as $l\rightarrow0$ and completely removes the presence of a collapse instability. This energetic behaviour is
demonstrated by the landscapes shown in
Fig.~\ref{fig:aF_vs_NF_vary_wB}(b).
\begin{list}{\labelitemi}{\leftmargin=1em}
\item{(i) Below the relevant soliton band (e.g. the band shaded with vertical hatch in Fig.~\ref{fig:aF_vs_NF_vary_wB}(a)) a significant collapse region is present and dominates the
energy landscape.  However, a positive region is just visible close to the origin in Fig.~\ref{fig:aF_vs_NF_vary_wB}(b)(i). }
\item{(ii) Solutions become supported when the
fermion-fermion interaction becomes larger and the play-off between all of the energy contributions generates a local minimum in the energy landscape (white cross).}
\item{(iii)  Above the soliton
band, the repulsive fermion-fermion interaction becomes so large
that it makes the system fully dispersive.}
\end{list}

Case (ii) is an intriguing prediction.  It suggests that the presence of repulsive {\it p}-wave fermion interactions leads to solitons which are the {\em global} energy minimum of the system.  This indicates that such solitons would be far more robust and stable than their bosonic counterparts, which are well-known to exist as meta-stable states prone to an irremovable collapse instability.

In Fig.~\ref{fig:aF_vs_NF_vary_NB} we show how the bands change with the number of bosons.  As the number of bosons decreases, the bands shift to lower $N_{\rm F}$ and become narrower.  Indeed, the bands scale approximately as $1/N_{\rm B}$, i.e. if we plot $N_{\rm F}/N_{\rm B}$ on the {\it x}-axis, the bands approximately overlap with each other.

\subsection{Soliton bands in $a_{\rm F}-a_{{\rm BF}s}$ space}
\begin{figure}[t]
\includegraphics[width=7cm,clip=true]{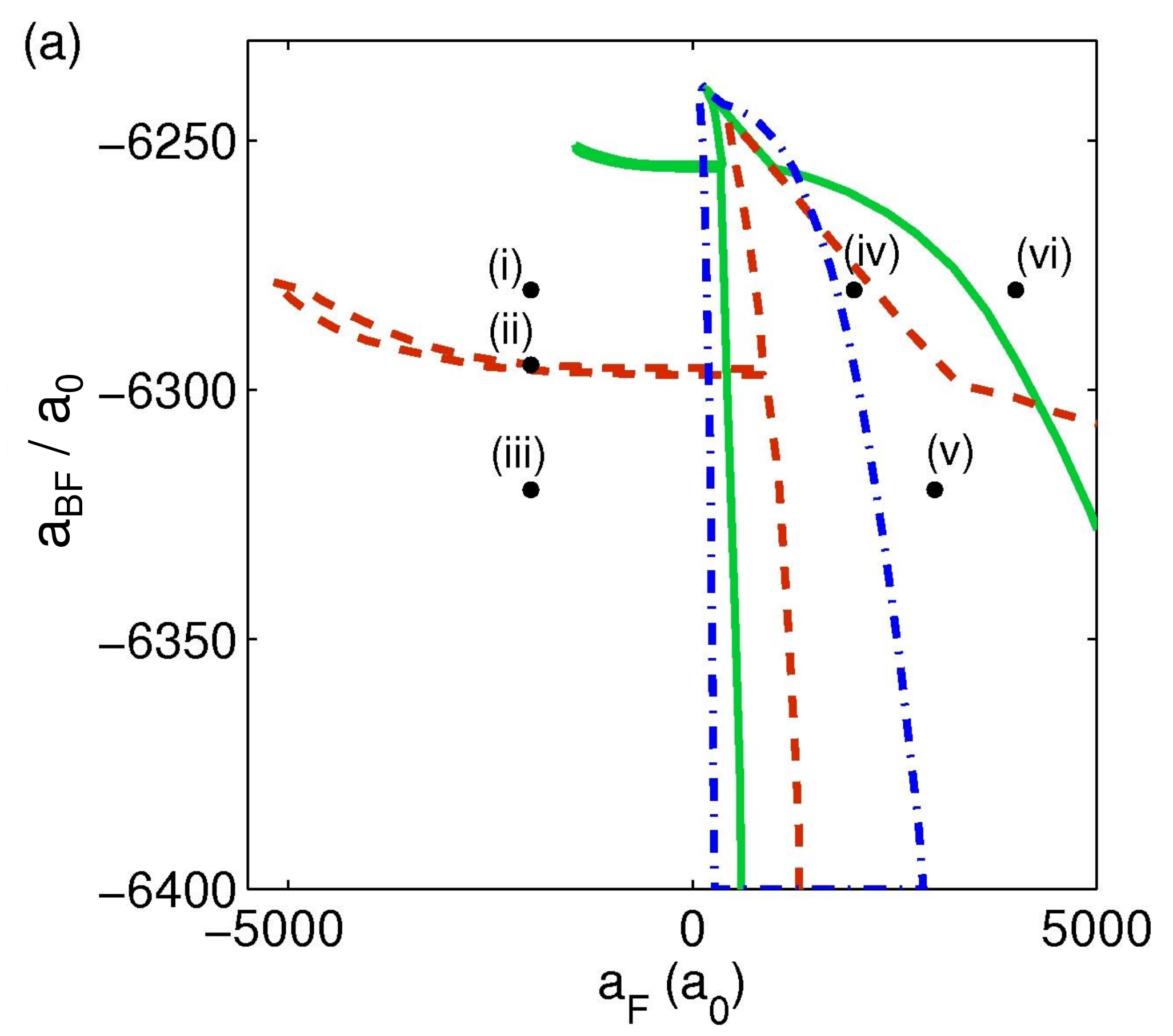}\\
\vspace{0.2cm}
\includegraphics[width=8cm, clip=true]{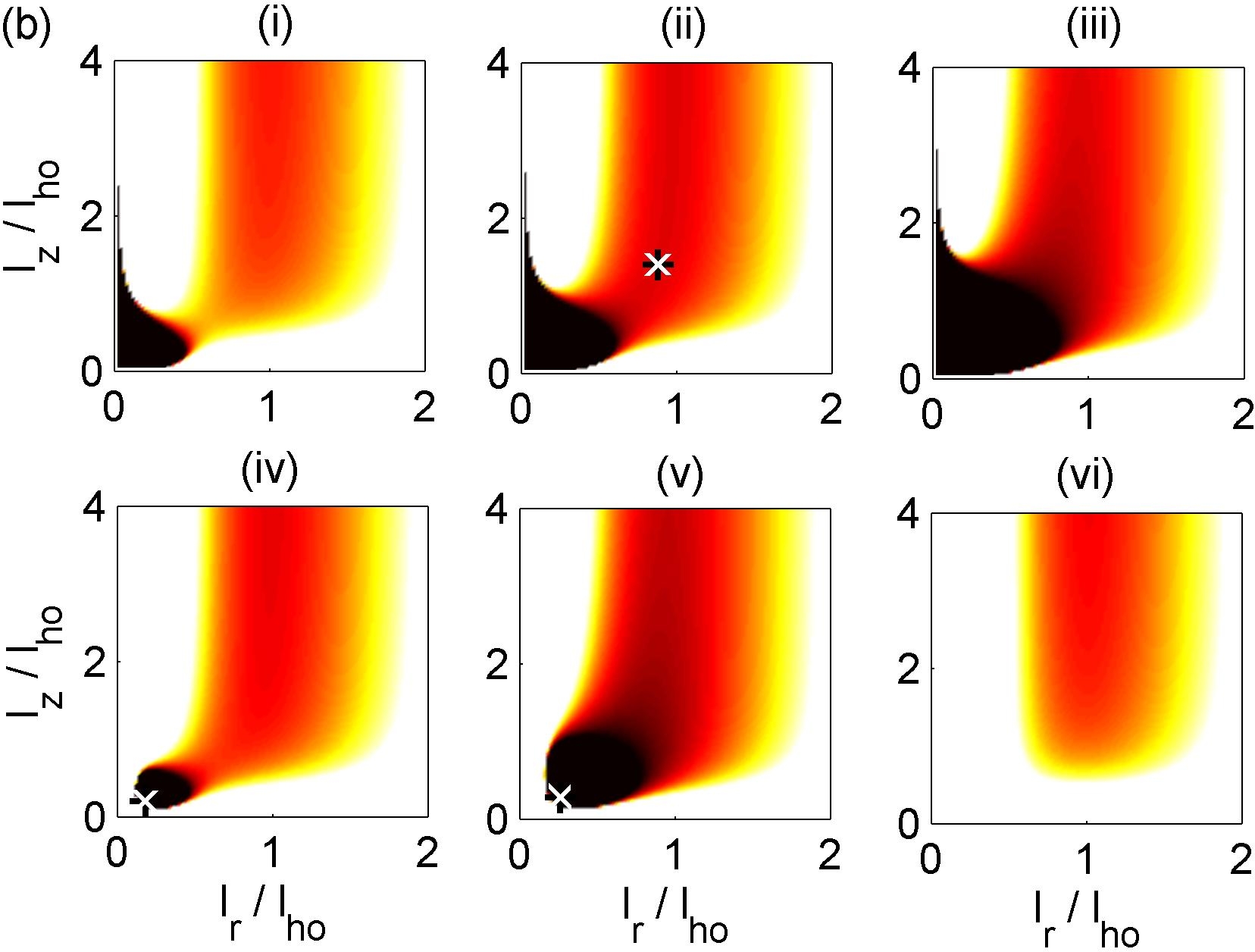}
\caption{Soliton bands (as shown by their boundary lines) in the parameter space of $a_{\rm F}$ and
$a_{{\rm BF}s}$. (a) We fix $N_{\rm B} =100,000$ and $N_{\rm F}=500$, and use $\omega_{\rm
B}/2\pi=10$ (red dashed), $100$ (green solid) and $1,000$ (blue dot-dashed) Hz. (b)
Energy landscapes for the $\omega_{\rm
B}/2\pi=10$ case at the points (i) $a_{\rm F}=-2000 a_0$, $a_{{\rm BF}s}=-6280a_0$, (ii) $a_{\rm F}=-2000 a_0$, $a_{{\rm BF}s}=-6295a_0$, (iii) $a_{\rm F}=-2000 a_0$, $a_{{\rm BF}s}=-6320a_0$, (iv) $a_{\rm F}=2000 a_0$, $a_{{\rm BF}s}=-6280a_0$, (v) $a_{\rm F}=3000 a_0$, $a_{{\rm BF}s}=-6320a_0$, and (iv) $a_{\rm F}=4000 a_0$, $a_{{\rm BF}s}=-6280a_0$.  These points are denoted in (a).}
\label{fig:aF_vs_aBF_vary_wB}
\end{figure}

The simultaneous manipulation of more than one scattering length has not been experimentally demonstrated.  However, in principle, this could be possible through a combination of magnetic, optical and confinement resonances.  With this is mind and by way of exploring the soliton solutions further, we now turn to examine the regions of soliton solutions in $a_{\rm F}-a_{{\rm
BF}s}$ space [Fig.~\ref{fig:aF_vs_aBF_vary_wB}(a)], for fixed $N_{\rm B}$ and $N_{\rm F}$.  We will see that parameter space is particularly interesting because it offers the possibility of forming bright soliton solutions as clear {\it global} minima. In this case, we observe complex-shaped regions of soliton solutions.  This includes a large cusp shaped region in the $a_{\rm F}>0$ half plane.  Although not shown, this region extends to indefinitely negative $a_{{\rm BF}s}$.  Furthermore, the soliton region features narrow `fingers' which extend far into the $a_{\rm F}<0$ half-plane.

We discriminate six distinct regions in this parameter space which we interpret by reference to their typical energy landscapes presented in Fig.~\ref{fig:aF_vs_aBF_vary_wB}(b):

\begin{list}{\labelitemi}{\leftmargin=1em}
\item{(i) Above this narrow band the landscape is
dominated by dispersion with a localised collapse region, but no local minimum exists.}
\item{(ii) Within the soliton band there exists a shallow energy minimum (case(ii)) adjacent to the collapse and dispersive regions.}
\item{(iii) Below the soliton band the whole landscape is unstable to collapse.}
\item{(iv, v) In the regions containing points (iv) and (v) there is no collapse region at the origin and there exists a well-localised and deep energy minimum denoting a soliton solution. }
\item{(vi) This region is dispersive due to the dominance of repulsive interactions.}
\end{list}

Regions (i),(ii),(iii) and (vi) possess energy landscapes which are analogous to those seen in Fig.~\ref{fig:vary_omega}(b). However, the most intriguing regions are (iv) and (v).  These solutions are the most common type that exist in this parameter space.  Like the observation in Fig.~\ref{fig:aF_vs_NF_vary_wB}(b)(ii), the soliton now becomes the {\em global} energy minimum of the system.  However, these landscapes are strikingly well localised and deep.  The depth of this minimum is typically of the order of $100 \hbar \omega_{\rm B}$.  For comparison, for a bright bosonic soliton the depth of the energy minima is of the order of $0.1 \hbar \omega_{\rm B}$.  Within the context of our scaling solutions (fixed gaussian shape), this depth and narrowness of the energy minima indicates extreme stability of the solutions to shape modification, including collapse and dispersion.
\begin{figure}[t]
\includegraphics[width=8.5cm,clip=true]{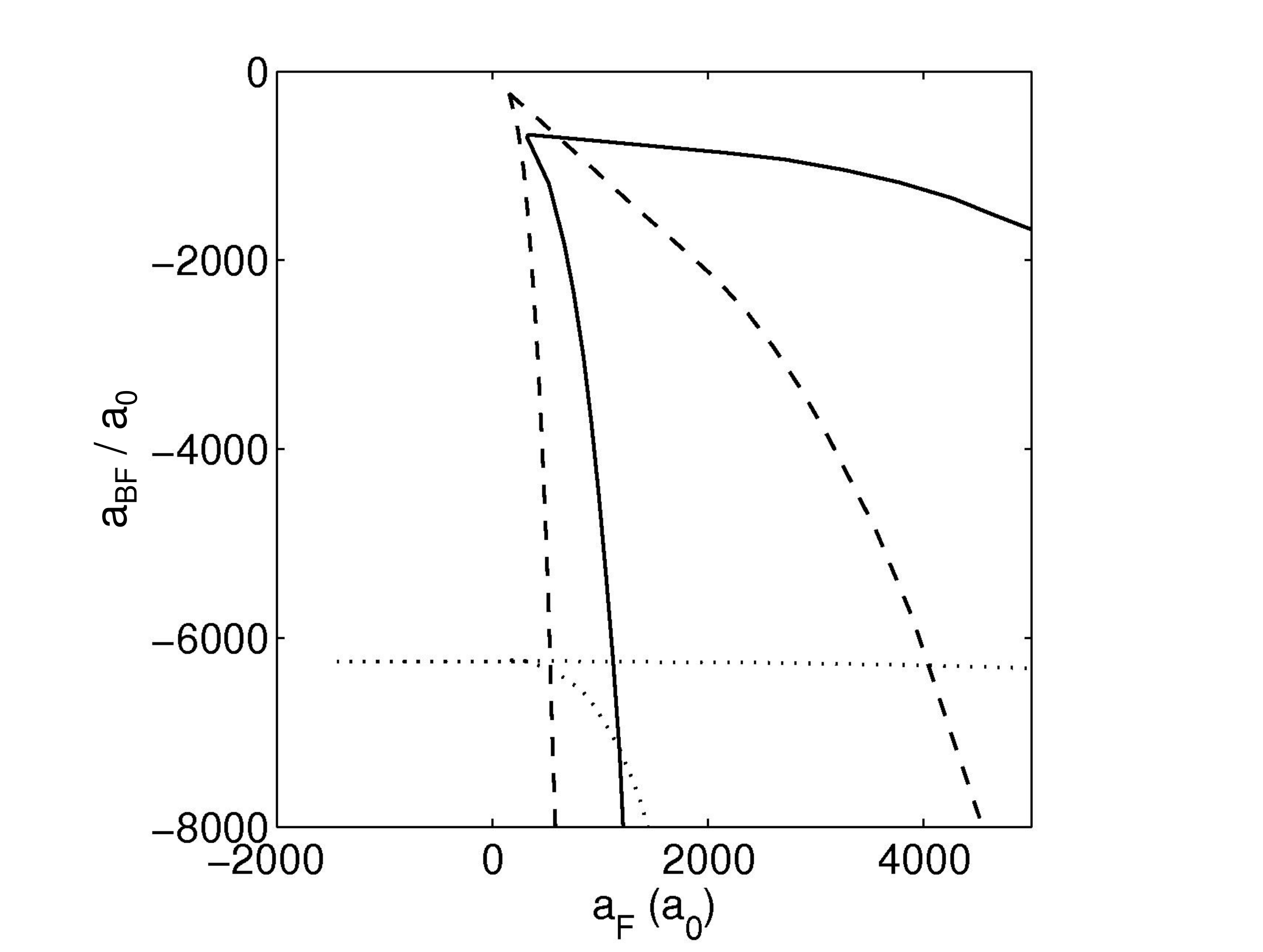}
\caption{Soliton bands in the parameter space $a_{\rm F}-a_{{\rm BF}s}$ (shown by their boundary lines). We fix $\omega_{\rm B}/2\pi=100$Hz and $N_{\rm F}=500$ and consider $N_{\rm B}=1,000$ (dashed line), $10,000$ (solid line) and $100,000$ (dotted line). }
\label{fig:aBF_vs_aF_vary_NB}
\end{figure}

For completenesss, Fig.~\ref{fig:aBF_vs_aF_vary_NB}
demonstrates how these extensive soliton regions in $a_{\rm F}-a_{{\rm BF}s}$ space become modified for different $N_B$.  The presence of the "fingers" is sensitive to $N_{\rm B}$ but the main region of solutions persists, albeit shifting to more negative $a_{{\rm BF}s}$ with increasing $N_{\rm B}$.

While here we have limited our study to the {\it p}-wave
interactions of only the fermions, the same qualitative soliton regimes and solutions
are obtained if the {\it p}-wave boson-fermion interaction is
included instead (or, indeed, if both are included).  This is
because the {\it p}-wave fermion-boson interaction has the same
functional form (scaling as $~l^{-5}$).

\section{Discussion and conclusion}
Our results show that in the absence of {\it p}-wave contributions
the range of soliton solutions is rather limited, being confined to
narrow bands in $a_{\rm BF}-N_{\rm F}$ space.  Indeed, the solutions behave similarly to bright bosonic solitons but with an effective scattering length.  While reducing the
radial confinement widens the soliton bands, this also makes the
system more prone to collective excitations which may disrupt the soliton.  For
a $^{87}$Rb-$^{40}$K mixture and in the validity regime of our approach ($N_{\rm B}, N_{\rm F}\gg1$) we cannot locate soliton solutions at
the native boson-fermion scattering length of $-215a_0$.  Values of around $-1000a_0$ are required to reach a soliton band at these atom numbers.  The work of Karpiuk {\it et al.} demonstrated that atom numbers of order unity are required to support Bose-Fermi solitons at the native $^{87}$Rb-$^{40}$K scattering length.  While the scattering length can be engineered to large values through scattering resonance tuning, the ensuing soliton bands are so narrow that experimentally locating them is likely to be problematic.
Furthermore, there remains a collapse instability in the system and the ratio of bosons to fermions is
constrained to small values.  In contrast, the presence of
repulsive {\it p}-wave fermion-fermion interactions has a dramatic
stabilizing effect on the system.  This can lead to a removal of the collapse instability such that the soliton solutions become {\em global} energy minima.  We find extensive soliton regimes, in which the soliton
minima are extremely deep, suggesting that they may form soliton
structures that are considerably more robust than in the absence of
{\it p}-wave interactions.  The {\it p}-wave interaction also
provides a strong tuning parameter, enabling the boson-fermion ratio
to be dramatically varied. Importantly, for a potential experimental
realization of BF solitons, we find extensive soliton solutions at
the native boson-fermion interaction and with only moderate
fermionic interactions.

The remarkable capacity of repulsive {\it p}-wave interactions to remove the collapse instability stems from the scaling behaviour of its energy contribution.  Denoting a generalized lengthscale of the gaussian wavepacket as $\ell$, the total variational energy of the Bose-Fermi system is of the form,
\begin{equation}
E\sim \frac{1}{\ell^2}+\ell^2 \pm \frac{1}{\ell^3}\pm \frac{1}{\ell^5}.
\end{equation}
The first two terms, the kinetic and potential energy terms, are always positive. The last two terms, the {\it s}-wave and {\it p}-wave interaction energies, respectively, may be positive or negative.  In the absence of {\it p}-wave interactions, a negative {\it s}-wave term will cause the energy to diverge to $-\infty$ as $\ell \rightarrow 0$, signifying the presence of the collapse instability.  For non-zero {\it p}-wave interactions, the {\it p}-wave term  dictates the fate of the system as $\ell \rightarrow 0$ and importantly, for positive {\it p}-wave interactions the collapse divergence is completely removed.

It is important to note that this scaling behaviour originates from the Thomas-Fermi approximation and so is not limited to gaussian wavepackets.  Consider homogeneous Bose and Fermi gases in a large hard-wall box of volume $\ell^3$.  It is trivial to see from Eqs. (\ref{eqn:GP_energy_functional}) and (\ref{eqn:fermion_energy_density}) that the energy scales as above, minus the $\ell^2$ term.  Thus it is clear that the capacity of repulsive {\it p}-wave interactions to stabilise against collapse is likely to extend to trapped Bose-Fermi mixtures in general.

While we have presented results for {\it p}-wave interactions in only the fermion-fermion case, we find qualitatively similar soliton regions, landscapes and conclusions when including boson-fermion {\it p}-wave interactions instead. This is because the same energy scaling discussed above applies.

In conclusion, according to a variational model, the presence of repulsive {\it p}-wave interactions in Bose-Fermi mixtures removes the usual collapse instability and leads to stable, robust bright soliton solutions that are global energy minima of the system.  We have discussed specifically the boson-fermion pairing of $^{87}$Rb and $^{40}$K, but the stabilizing effect of repulsive {\it p}-wave interactions should apply in general to other ultracold Bose-Fermi mixtures.  Given that the collapse instability has proved a major hindrance to the controlled generation, manipulation and interaction of matter-wave solitons to date, these more stable {\it p}-wave entities may provide a more versatile route to explore and exploit the special characteristics of solitons.

\end{document}